\documentclass[prl,reprint,superscriptaddress,preprintnumbers,nofootinbib]{revtex4-2}
\usepackage{lineno}
\usepackage{comment}
\usepackage{amsmath}
\usepackage{graphicx} 
\usepackage{ulem}
\usepackage{dcolumn} 
\usepackage{bm} 
\usepackage{hyperref}
\hypersetup{colorlinks=true,citecolor=blue, linkcolor=blue}
\usepackage[range-phrase={\,\text{--}\,}, range-units={single}, separate-uncertainty=true]{siunitx}

\usepackage{soul}   
\setstcolor{red}    
\usepackage[usenames, dvipsnames]{xcolor} 

\DeclareSIUnit\year{yr}
\DeclareSIUnit\solarMass{\ensuremath{\mathit{M}_\odot}}

\begin{document}
	
	\title{First measurement of the low-energy direct capture in $\mathrm{^{20}Ne(p,\gamma)^{21}Na}$ and improved energy and strength of the $E^{\rm }_{cm} = 368$~keV resonance}
	
	\author{E.~Masha} \affiliation{Helmholtz-Zentrum Dresden-Rossendorf, 01328 Dresden, Germany} \affiliation{Universit\`a degli Studi di Milano, 20133 Milano, Italy}\affiliation{INFN, Sezione di Milano, 20133 Milano, Italy}
	
	\author{L.~Barbieri}\affiliation{Universit\`a degli Studi di Padova, 35131 Padova, Italy}
	\affiliation{INFN, Sezione di Padova, 35131 Padova, Italy}
	
	\author{J.~Skowronski}\affiliation{Universit\`a degli Studi di Padova, 35131 Padova, Italy}
	\affiliation{INFN, Sezione di Padova, 35131 Padova, Italy}
	
	\author{M.~Aliotta}\affiliation{SUPA, School of Physics and Astronomy, University of Edinburgh, EH9 3FD Edinburgh, United Kingdom}
	
	\author{C.~Ananna}\affiliation{Universit\`a degli Studi di Napoli “Federico II”, 80126 Napoli, Italy} \affiliation{INFN, Sezione di Napoli, 80126 Napoli, Italy}
	
	\author{F.~Barile}\affiliation{Universit\`a degli Studi di Bari, 70125 Bari,  Italy}
	\affiliation{INFN, Sezione di Bari, 70125 Bari, Italy}
	
	\author{D.~Bemmerer}\affiliation{Helmholtz-Zentrum Dresden-Rossendorf, 01328 Dresden, Germany}
	
	\author{A.~Best}\affiliation{Universit\`a degli Studi di Napoli “Federico II”, 80126 Napoli, Italy} \affiliation{INFN, Sezione di Napoli, 80126 Napoli, Italy}
	
	\author{A.~Boeltzig}\affiliation{Helmholtz-Zentrum Dresden-Rossendorf, 01328 Dresden, Germany}
	
	\author{C.~Broggini}\affiliation{INFN, Sezione di Padova, 35131 Padova, Italy}
	
	\author{C.G.~Bruno}\affiliation{SUPA, School of Physics and Astronomy, University of Edinburgh, EH9 3FD Edinburgh, United Kingdom}
	
	\author{A.~Caciolli}\email{antonio.caciolli@unipd.it}\affiliation{Universit\`a degli Studi di Padova, 35131 Padova, Italy} \affiliation{INFN, Sezione di Padova, 35131 Padova, Italy}
	
	\author{M.~Campostrini}\affiliation{Laboratori Nazionali di Legnaro, 35020, Legnaro (PD), Italy}

 \author{F.~Casaburo} \affiliation{Universit\`a degli Studi di Genova, 16146 Genova, Italy}
	\affiliation{INFN, Sezione di Genova, 16146 Genova, Italy}
 
	\author{F.~Cavanna}\affiliation{INFN, Sezione di Torino, 10125 Torino, Italy}
	
	\author{G.F.~Ciani}\affiliation{Universit\`a degli Studi di Bari, 70125 Bari,  Italy}
	\affiliation{INFN, Sezione di Bari, 70125 Bari, Italy}
	
	\author{A.~Ciapponi}\affiliation{Universit\`a degli Studi di Milano, 20133 Milano, Italy} \affiliation{INFN, Sezione di Milano, 20133 Milano, Italy}
	
	\author{P.~Colombetti} \affiliation{Universit\`a degli Studi di Torino, 10125 Torino, Italy} \affiliation{INFN, Sezione di Torino, 10125 Torino, Italy}
	
	\author{A.~Compagnucci}\affiliation{Gran Sasso Science Institute, 67100 L'Aquila, Italy}
	\affiliation{INFN, Laboratori Nazionali del Gran Sasso (LNGS), 67100 Assergi, Italy}
	
	\author{P.~Corvisiero}\affiliation{Universit\`a degli Studi di Genova, 16146 Genova, Italy}
	\affiliation{INFN, Sezione di Genova, 16146 Genova, Italy}
	
	\author{L.~Csedreki}\affiliation{Institute for Nuclear Research (Atomki), PO Box 51, 4001 Debrecen, Hungary}
	
	\author{T.~Davinson}\affiliation{SUPA, School of Physics and Astronomy, University of Edinburgh, EH9 3FD Edinburgh, United Kingdom}
	
	\author{R.~Depalo}\affiliation{Universit\`a degli Studi di Milano, 20133 Milano, Italy} \affiliation{INFN, Sezione di Milano, 20133 Milano, Italy}
	
	\author{A.~Di~Leva}\affiliation{Universit\`a degli Studi di Napoli “Federico II”, 80126 Napoli, Italy} \affiliation{INFN, Sezione di Napoli, 80126 Napoli, Italy}
	
	\author{Z.~Elekes}\affiliation{Institute for Nuclear Research (Atomki), PO Box 51, 4001 Debrecen, Hungary}
	
	\author{F.~Ferraro}\affiliation{Universit\`a degli Studi di Milano, 20133 Milano, Italy} \affiliation{INFN, Sezione di Milano, 20133 Milano, Italy}
	
	\author{E.M.~Fiore}\affiliation{Universit\`a degli Studi di Bari, 70125 Bari, Italy}
	\affiliation{INFN, Sezione di Bari, 70125 Bari, Italy}
	
	\author{A.~Formicola}\affiliation{INFN, Sezione di Roma, 00185 Roma, Italy}
	
	\author{Zs.~F\"ul\"op}\affiliation{Institute for Nuclear Research (Atomki), PO Box 51, 4001 Debrecen, Hungary}
	
	\author{G.~Gervino} \affiliation{Universit\`a degli Studi di Torino, 10125 Torino, Italy}\affiliation{INFN, Sezione di Torino, 10125 Torino, Italy}
	
	\author{A.~Guglielmetti}\affiliation{Universit\`a degli Studi di Milano, 20133 Milano, Italy} \affiliation{INFN, Sezione di Milano, 20133 Milano, Italy}
	
	\author{C.~Gustavino}\affiliation{INFN, Sezione di Roma, 00185 Roma, Italy}
	
	\author{Gy.~Gy\"urky}\affiliation{Institute for Nuclear Research (Atomki), PO Box 51, 4001 Debrecen, Hungary}
	
	\author{G.~Imbriani}\affiliation{Universit\`a degli Studi di Napoli “Federico II”, 80126 Napoli, Italy} \affiliation{INFN, Sezione di Napoli, 80126 Napoli, Italy}
    \author{J.~Jos\'e}\affiliation{Universitat Politècnica de Catalunya, 08019 Barcelona, Spain}
    \affiliation{Institut d'Estudis Espacials de Catalunya, 08034 Barcelona, Spain}
	\author{M.~Junker}\affiliation{INFN, Laboratori Nazionali del Gran Sasso (LNGS), 67100 Assergi, Italy}	
	\author{M.~Lugaro}\affiliation{Konkoly Observatory, Research Centre for Astronomy and Earth Sciences, HUN-REN, 1121 Budapest, Hungary}
	\affiliation{ELTE E\"{o}tv\"{o}s Lor\'and University, Institute of Physics, 1117 Budapest, Hungary}

 \author{P.~Manoj}\affiliation{INFN, Sezione di Torino, Via P. Giuria 1, 10125 Torino, Italy}
 
	\author{P.~Marigo}\affiliation{Universit\`a degli Studi di Padova, 35131 Padova, Italy}
	\affiliation{INFN, Sezione di Padova, 35131 Padova, Italy}
	
	\author{R.~Menegazzo}\affiliation{INFN, Sezione di Padova, 35131 Padova, Italy}
	
	\author{V.~Paticchio}\affiliation{INFN, Sezione di Bari, 70125 Bari, Italy}
	
	\author{D.~Piatti}\affiliation{Universit\`a degli Studi di Padova, 35131 Padova, Italy}
	\affiliation{INFN, Sezione di Padova, 35131 Padova, Italy}
	
	\author{P.~Prati}\affiliation{Universit\`a degli Studi di Genova, 16146 Genova, Italy}
	\affiliation{INFN, Sezione di Genova, 16146 Genova, Italy}

	\author{D.~Rapagnani}\affiliation{Universit\`a degli Studi di Napoli “Federico II”, 80126 Napoli, Italy} \affiliation{INFN, Sezione di Napoli, 80126 Napoli, Italy}
	
	\author{V.~Rigato}\affiliation{Laboratori Nazionali di Legnaro, 35020, Legnaro (PD), Italy}
	\author{D.~Robb}\affiliation{SUPA, School of Physics and Astronomy, University of Edinburgh, EH9 3FD Edinburgh, United Kingdom}
	\author{L.~Schiavulli}\affiliation{Universit\`a degli Studi di Bari, 70125 Bari, Italy}
	\affiliation{INFN, Sezione di Bari, 70125 Bari, Italy}
	
	\author{R.S.~Sidhu}\affiliation{SUPA, School of Physics and Astronomy, University of Edinburgh, EH9 3FD Edinburgh, United Kingdom}
	
	\author{O.~Straniero}\affiliation{INAF Osservatorio Astronomico d'Abruzzo, 64100 Teramo, Italy}\affiliation{INFN, Sezione di Roma, 00185 Roma, Italy}
	
	\author{T.~Sz\"ucs}\affiliation{Institute for Nuclear Research (Atomki), PO Box 51, 4001 Debrecen, Hungary}
	
	\author{S.~Zavatarelli}\email{sandra.zavatarelli@ge.infn.it}\affiliation{Universit\`a degli Studi di Genova, 16146 Genova, Italy} \affiliation{INFN, Sezione di Genova, 16146 Genova, Italy}
	
	\collaboration{LUNA collaboration}
	
	\date{\today}
	
	\begin{abstract}
The $\mathrm{^{20}Ne(p, \gamma)^{21}Na}$ reaction is the slowest in the NeNa cycle and directly affects the abundances of the Ne and Na isotopes in a variety of astrophysical sites.
Here we report the measurement of its direct capture contribution, for the first time below $E\rm_{cm} = 352$~keV, and of the contribution from the $E^{\rm }_{cm} = 368$~keV resonance, which dominates the reaction rate at $T=0.03-1.00$~GK.
The experiment was performed deep underground at the Laboratory for Underground Nuclear Astrophysics, using a high-intensity proton beam and a windowless neon gas target.
Prompt $\gamma$ rays from the reaction were detected with two high-purity germanium detectors.
We obtain a resonance strength $\omega \gamma~=~(0.112 \pm 0.002_{\rm stat}~\pm~0.005_{\rm sys})$~meV, with an uncertainty a factor of $3$ smaller than previous values.
Our revised reaction rate is 20\% lower than previously adopted at $T < 0.1$~GK and agrees with previous estimates at temperatures $T \geq 0.1$~GK.
		Initial astrophysical implications are presented.
	\end{abstract}
	\maketitle
	The NeNa cycle converts hydrogen into helium using neon and sodium isotopes as catalysts through the following reactions: 
	\begin{gather*}
		\mathrm{^{20}Ne(p,\gamma)^{21}Na(\beta^{+}\nu)^{21}Ne(p,\gamma)^{22}Na(\beta^{+}\nu)}\\
		\mathrm{^{22}Ne(p,\gamma)^{23}Na(p, \alpha)^{20}Ne}.\label{eq:neonprod}
	\end{gather*}
	
	The ashes of this nucleosynthesis sequence may become visible when they are carried to the stellar surface as a consequence of mixing with the stellar interior.
	
	Mixing occurs, for example, in asymptotic giant branch (AGB) stars of masses $M=5-9 M_\odot$, where the convective envelope reaches into the H-burning layers, bringing freshly synthesised material to the stellar surface, a phenomenon known as Hot Bottom Burning (HBB) \cite{renzinivoli_1981, Ventura_2011, ventura_2014}. 
	As a result, the atmosphere of these massive AGB stars becomes enriched in nitrogen and sodium.
	Another astrophysical object that is affected by the NeNa cycle is ONe novae.
	
	In particular, the  $1275$~keV $\gamma$-ray line associated with the $\beta^+$ decay of $^{22}$Na, would be essential to confirm a long-lasting prediction of nova nucleosynthesis models \cite{Jose16-Book}.
	
\begin{figure*}[ht]
		\centering
		\includegraphics[width=0.8\textwidth]{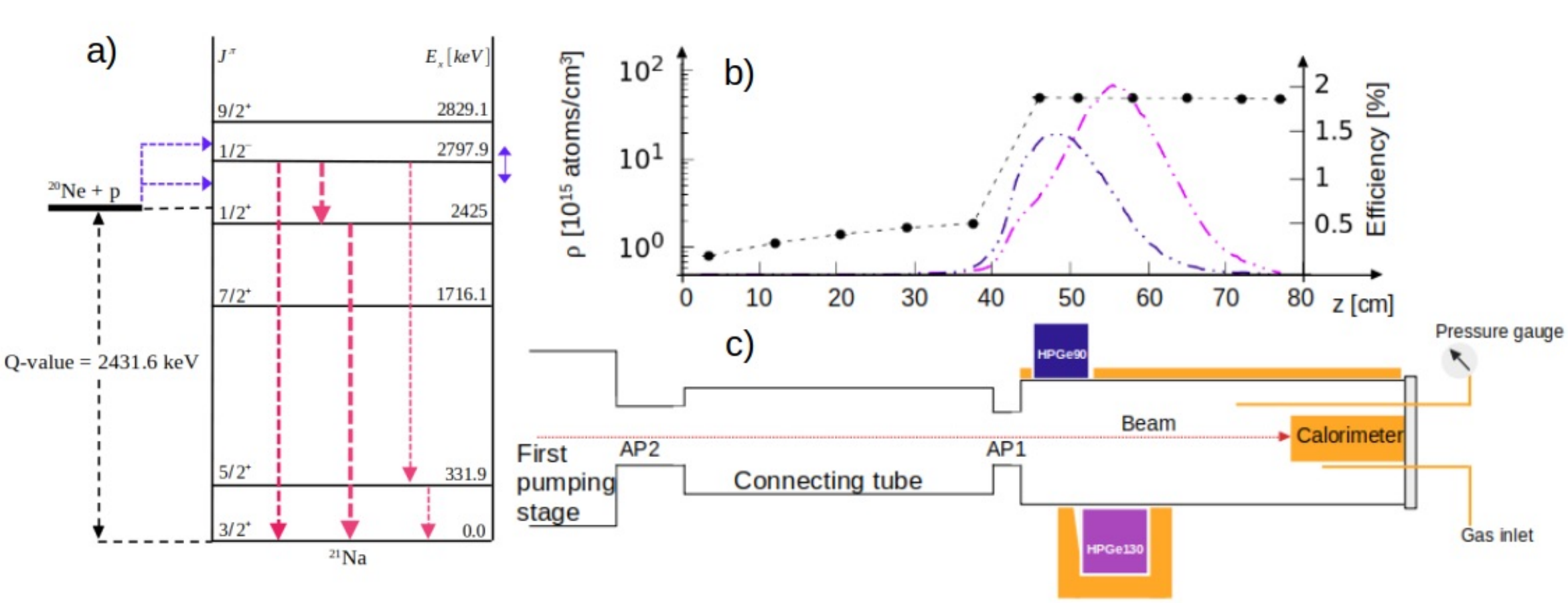}
		\caption{a): Level scheme of $\mathrm{^{21}Na}$. The transitions from the $E^{\rm }_{cm} \simeq 366$~keV resonance are shown in red while the blue arrow indicates the direct capture energy range explored in this work.
			b): Gas density profile (black data points) along the beam line through apertures AP2 and AP1 and into the gas chamber. Blue and magenta dash-dotted lines show the detection efficiencies (right $y$-axis) for detectors HPGe90 and HPGe130, respectively, along the beam axis, as a function of distance $z$ from AP2.  
			The efficiency curves refer to the $373$~keV $\gamma$ ray emitted by the $\mathrm{^{20}Ne(p,\gamma)^{21}Na}$ reaction. 
			c): Simplified sketch of the experimental setup used for the study of the $\mathrm{^{20}Ne(p, \gamma)^{21}Na}$ reaction at LUNA. The beam enters the chamber from the left through the last collimator (AP1) and stops onto a copper calorimeter. 
		}
		\label{fig_1}
	\end{figure*}

	The $^{20}$Ne is the most abundant isotope of those participating to the NeNa cycle, with the $^{20}$Ne(p,$\gamma$)$^{21}$Na being the slowest reaction in the cycle and thus 
	affecting the final abundances of the Ne and Na isotopes.
	A sensitivity study on the effect of a variation in the $^{20}$Ne(p,$\gamma$)$^{21}$Na rate on novae ejecta, suggests a significant impact on isotopic abundances of elements with $A<40$  \cite{Iliadis02-ApJSS}.
	While other reactions in the cycle are now well constrained following recent measurements of the $\mathrm{^{22}Ne(p,\gamma)^{23}Na}$, \cite{Cavanna15-PRL, Ferraro18-PRL, Kelly17-PRC, William20}, the $\mathrm{^{20}Ne(p,\gamma)^{21}Na}$ 
	and the $\mathrm{^{23}Na(p, \alpha)^{20}Ne}$ reactions, the first and the last of the NeNa cycle, are still carrying the largest uncertainties.
 Here we focus on the $\mathrm{^{20}Ne(p,\gamma)^{21}Na}$ reaction.
 At temperatures $T< 0.1$~GK, relevant for HBB, the $\mathrm{^{20}Ne(p,\gamma)^{21}Na}$ reaction ($Q$-value = $2431.9$~keV) is dominated by the high energy tail of a sub-threshold state at $E\mathrm{_{cm}}$ = $\mathrm{-}$$6.7$~keV ($\Gamma\mathrm{_{\gamma}}$ = $0.31 \pm 0.07$~eV \cite{Rolfs75-NPA}), corresponding to the $E\mathrm{_{x}}$ = $2425$~keV excited level in $^{21}$Na \cite{Lyons18-PRC}. 
	At temperatures $T = 0.1 - 1.0$~GK, including those relevant to novae, the rate is governed, instead, by a narrow resonance at $E^{\rm }_{cm} \simeq 366$~keV \cite{Rolfs75-NPA}, corresponding to an excited state at $E\mathrm{_{x}} = 2799$~keV in $\mathrm{^{21}Na}$ (Fig.~\ref{fig_1}a), and by direct capture contributions to the ground-, first-, and second excited states in $^{21}$Na at $E\mathrm{_{x}} = 332$ and $2425$~keV, respectively.
	The strength of the narrow resonance at $E^{\rm }_{cm} \simeq 366$~keV was measured for the first time by Rolfs $et$ $al.$ \cite{Rolfs75-NPA} to be $\omega \gamma$ = ($0.11 \pm 0.02$)~meV and by a recent study (Cooper, PhD thesis, \cite{Cooper-PhD}), which instead reports a strength of $\omega \gamma = ( 0.0722 \pm 0.0068$) meV.
	
	Direct capture contributions at $E\mathrm{_{cm}} \geq 352$~keV, as well as contributions from higher-energy resonances, have also been reported in previous works \cite{tanner59, VANDERLEUN1964, Rolfs75-NPA, Keinonen77-PRC, Mukhamedzhanov2006AsymptoticNC, Christian_2013, Lyons18-PRC}.
	Specifically, a non-resonant component was first investigated in Ref. \cite{tanner59} at beam energies $E\mathrm{_{cm}}$ = $600$~keV and $E\mathrm{_{cm}}$ = 1050 keV, using the activation method, i.e., exploiting the $\mathrm{\beta^{+}}$-decay of $\mathrm{^{21}Na}$ (half-life $t\mathrm{_{1/2}} = 22.4$~s \cite{Shidling_22}) into $\mathrm{^{21}Ne}$. 
	The subsequent comprehensive study by Rolfs $et$ $al.$ \cite{Rolfs75-NPA} investigated the direct component and several resonances at proton beam energies  $E_{\rm {cm}} = 352 - 2000$~keV.
	The direct capture into the $2425$~keV state was found to be dominant \cite{Rolfs75-NPA}. 
	More recently, the $\mathrm{^{20}Ne(p,\gamma)^{21}Na}$ reaction was studied indirectly using the $\mathrm{^{20}Ne(^{3}He,d)^{21}Na}$ reaction \cite{Mukhamedzhanov2006AsymptoticNC}. 
	The partial width of the sub-threshold state and the direct capture spectroscopic factors were calculated using the asymptotic normalization coefficient (ANC) formalism \cite{Mukhamedzhanov2006AsymptoticNC}. 
	The results are in good agreement with previous data \cite{Rolfs75-NPA} for the direct capture to the $2425$~keV sub-threshold state, while a discrepancy of 65\% was found for the direct capture into the ground state.
	New direct capture data were also recently reported by Lyons \textit{et al.} \cite{Lyons18-PRC} at energies $E\mathrm{_{cm}}= 477 - 1905$~keV, and by Karpesky (PhD thesis) \cite{Karpesky-PhD} at energies $E\mathrm{_{p}} < 400$~keV. 
	In the latter study, the direct capture and resonant components could not be clearly distinguished, while the results for the direct capture to the ground state were found to be $\simeq 40$\% lower than those by Rolfs \textit{et al.}. \cite{Rolfs75-NPA}.
	As low energy data on the direct capture are either lacking or carrying high uncertainties and given that the two available data sets \cite{Rolfs75-NPA, Cooper-PhD} on the  $E^{\rm }_{cm} \simeq 366$~keV strengths are in disagreement, improved measurements are needed to better constrain the $\mathrm{^{20}Ne(p, \gamma)^{21}Na}$ reaction rate.
	
\begin{figure}[ht]
		\includegraphics[width=0.5\textwidth]{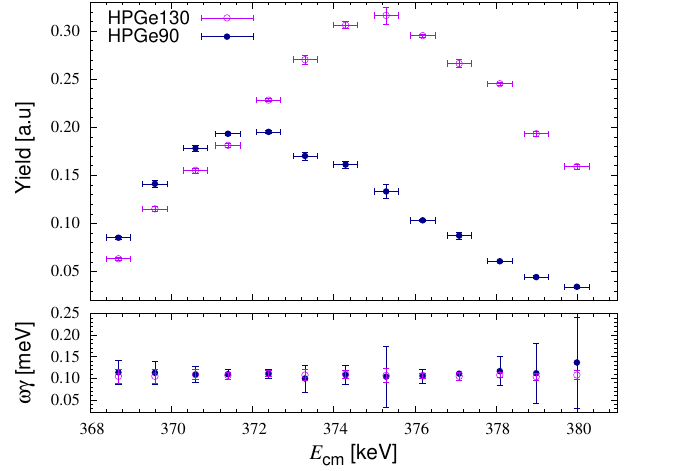}
		\caption{Top panel: Experimental yields for the $E\mathrm{_{\gamma}}$ = $2425$~keV as obtained with the HPGe130 (magenta) and HPGe90 (dark-blue) detectors as a function of proton beam energy. 
			Bottom panel: Resonance strength determined from experimental yields at each beam energy studied (see text for details).}
		\label{fig_2}
	\end{figure}

	Here, we report on the measurements  performed at the Laboratory for Underground Nuclear Astrophysics (LUNA) \cite{Aliotta-22} exploiting the low environmental background level \cite{Caciolli09-EPJA,Szucs10-EPJA} of the Gran Sasso National Laboratories (LNGS), Italy.
	The setup used was similar to that adopted for the study of the $\mathrm{^{22}Ne(p,\gamma)^{23}Na}$ reaction \cite{Cavanna15-PRL, Cavanna14-EPJA}. 
	A schematic view is shown in Fig.~\ref{fig_1}. 
	Briefly, an intense ($\sim 300$ $\mathrm{\mu }$A) proton beam from the LUNA 400~kV accelerator \cite{Formicola03-NIMA} was delivered onto a windowless chamber filled with neon gas of natural composition ($90.48\%$ $\mathrm{^{20}Ne}$, $0.27\%$ $\mathrm{^{21}Ne}$, 9.25$\%$ $\mathrm{^{22}Ne}$). 
	The gas was maintained within the windowless chamber by a differential pumping system, through three apertures of different diameters (see \cite{Mossa20-EPJA} for details).
	The beam entered the target chamber through aperture AP1 ($4$~cm in length and $7$~mm in diameter) and was stopped on a calorimeter for beam current measurement \cite{Ferraro18-EPJA}. 
	The prompt $\gamma$ rays from the 
	$\mathrm{^{20}Ne(p, \gamma)^{21}Na}$ reaction 
	were detected with two high purity germanium detectors of 90\% (HPGe90) and 130\% (HPGe130) relative efficiency, with faces centered at two different positions corresponding to a distance, respectively, of $5.6$~cm and $13.4$~cm from AP1 along the beam axis   (Fig.~\ref{fig_1}). 
	The HPGe130 was surrounded by a $4$~cm thick copper shielding and the entire setup (gas target and both detectors) was surrounded by $20 - 30$~cm thick lead bricks (not shown in the figure) to suppress the laboratory environmental background. 
	The entire lead castle was finally enclosed in a Plexiglas anti-radon box filled with an over-pressure of $\mathrm{N_{2}}$ to avoid radon gas inside the lead shielding.
	Unlike the setup adopted in Ref. \cite{Cavanna15-PRL}, neither HPGe detector was collimated in the present study. 
	As a result, detection efficiencies were maximized at different positions inside the target chamber as shown in the bottom panel of Fig.~\ref{fig_1} (right $y$-axis). 
	The pre-amplified signal from each detector was sent to an amplifier (ORTEC Spectroscopy Amplifier 672) and then acquired by an MCA-ADC (EtherNIM analog multichannel analyzer). 
	The dead-time for each detector was $\sim$ 1\% during all data taking.
 The $\gamma$-ray detection efficiency was measured at several positions along the beam axis (in $5$~mm steps) using point-like radioactive sources ($^{133}$Ba, $^{137}$Cs and $^{60}$Co), with activities calibrated by the Physikalisch-Technische Bundesanstalt (PTB \cite{PTB}) to 1\% accuracy.
 Efficiency measurements were extended to higher energies (up to $6.8$~MeV) using the well-known $^{14}$N(p,$\gamma$)$^{15}$O resonance at $E\mathrm{_{cm}}$ = $259$~keV \cite{Daigle16-PRC}. 
	The experimental setup (Fig.~\ref{fig_1}) was implemented in the LUNA GEANT code \cite{Mossa20-EPJA, Mossa20-EPJA}. 
	The geometry of the code was fine-tuned through a detailed comparison with experimental data obtained from the radioactive sources and the $^{14}$N(\textit{p},$\gamma$)$^{15}$O reaction.
	For the study of the resonant capture contribution, experimental yields were measured in the range $E_{\rm cm} = 366-380$~keV in $1-2$~keV steps.
	Yield profiles for both detectors are shown in Fig.~\ref{fig_2} for the strongest transition at $E_\gamma = 2425$~keV.  
	The energy of the resonance $E^{\rm }_r$ was determined by taking into account the energy loss of the beam in the gas target, as: 
	\begin{equation}
		E^{\rm res}_{cm}= E_\mathrm{{cm}} -  \left(\int_{z_{0}}^{z_{\rm max}} \frac{\mathrm{d}E_{p}}{\mathrm{d}(\rho z)}\rho(z) \mathrm{d}z \right)_{\rm cm},
		\label{eq_1}
	\end{equation}
	where $E\mathrm{^{\rm res}_{\rm cm}}$ is the resonance energy in the center-of-mass system, $E\mathrm{_{cm}}$ is the energy corresponding to the maximum of the yield profile, as obtained by a fit to experimental data; $\frac{\mathrm{d}E}{\mathrm{d}(\rho z)}$ is the stopping power of protons in neon gas, given by SRIM \cite{Ziegler10-NIMB}; $z_{0}$= 0 corresponds to the entrance position of the beam in the first pumping stage (Fig.~\ref{fig_1}); and $z_{\rm max}$ is the position at which the detection efficiency reaches its maximum, for a given detector.
	The density profile, $\rho(z)$, which affects the target thickness and therefore the resonance energy determination, has been corrected for the beam heating effect following the prescription in Ref. \cite{Cavanna14-EPJA}, where an identical setup was used.
	The energy of the resonance was calculated using Eq. \ref{eq_1} for each detector separately, leading to a weighted average of 
	$E\mathrm{_{cm}}$ = ($368.0 \pm 0.5$)~keV, consistent with the value $E\mathrm{_{cm}}$ = ($366 \pm 5$)~keV reported by Rolfs \textit{et al.} \cite{Rolfs75-NPA}.
	The overall uncertainty on the resonance energy, 10 times smaller than the literature value, is obtained from  error propagation of uncertainties on: beam energy ($0.3$~keV \cite{Formicola03-NIMA}), proton energy loss in the neon gas target (1.7$\%$ \cite{Ziegler10-NIMB}), proton energy corresponding to the maximum of the fitted yield profile ($< 0.2$~keV), and beam heating correction ($1.6\%$).
	For each beam energy, we also determined the branching ratios of all transitions de-exciting the $E\mathrm{_{cm}} = 368$~keV resonance to the $^{21}$Na ground-, the first-, and second excited states (${\rm R} \rightarrow {\rm GS}$, ${\rm R}  \rightarrow 332$~keV, ${\rm R}  \rightarrow 2425$~keV, respectively). 
	The branching ratios were obtained as the ratio between the efficiency-corrected yield of a single transition and the sum of all observed transitions.
	Weighted average values are given in Table \ref{tab_1}, together with literature values \cite{Rolfs75-NPA}.
	\begin{table}[h!!]
		\centering
		\caption{Branching ratios for transitions from the $E\mathrm{_{cm}} = 368.0$~keV resonance for   the present work (LUNA) and the literature  \cite{Rolfs75-NPA}. 
			Uncertainties in the LUNA branching ratios include  statistical and systematic contributions.}
		\begin{tabular}{lcc}
			\hline \hline
			Transition & LUNA & Rolfs \textit{et al.} \cite{Rolfs75-NPA}\\
			\hline
			${\rm R}  \rightarrow 2425$ & $57 \pm 2$ & $56 \pm 4$\\
			${\rm R}  \rightarrow  332$ & \; $4.0 \pm 0.2$ & $11 \pm 4$\\
			${\rm R} \rightarrow  {\rm GS}$ & $39 \pm 2$ & $33 \pm 4$\\
			\hline \hline
		\end{tabular}
		\label{tab_1}
	\end{table}
	
	Under the assumption of a thick-target yield condition \cite{Iliadis07-Book}, the resonance strength can be obtained directly from the experimental total yields $Y$ (i.e., summed over all transitions), as: 
	\begin{equation}
		\omega \gamma = \frac{2 Y}{\lambda_r^2}\epsilon_r \frac{M}{m+M},
		\label{yield_omega}
	\end{equation}
	where $\lambda_r$ is the de-Broglie wavelength at the resonance energy (in the center of mass system) and $\epsilon_r M/(m+M)$ is the effective stopping power in the center of mass system (with projectile and target masses $m$ and $M$, respectively).
	
	Since the resonance width ($\Gamma \simeq 5$~meV) is much smaller compared to the beam energy loss in the target ($\Delta E \simeq 15-20$~keV), the thick-target condition is satisfied at all beam energies investigated here, and the resonance is populated at different positions in the gas target, depending on beam energy.
	However, despite the narrow width of the resonance, the distribution of the emitted $\gamma$ rays along the $z$-axis is not point-like. 
	When the energetically narrow beam ($\Delta E\mathrm{_{beam}} \sim 0.1$ keV \cite{Formicola03-NIMA}) goes through the gas target, its energy distribution widens because of the straggling effect and the resonance condition is reached over a broader target region. 
	At the center of the target, the energy broadening of the beam is of the order of $1.5$~keV.  
	This effect has to be combined with the detection efficiency, as discussed in Ref. \cite{Bemmerer18-EPL}.  
	To correct experimental yields for different combinations of efficiency and beam straggling, simulations were performed using the LUNA GEANT code for each point of the yield curve and for each detector. 
	Corrected yields were then used in Eq. \ref{yield_omega} to arrive at individual $\omega \gamma$ values as shown in Fig. \ref{fig_2} (bottom panel). 
	Resonance strength values were found to agree within $1 \sigma$ of each other at all beam energies (i.e., at all positions in the target chamber), and led to a weighted average of $\omega \gamma = (0.112 \pm 0.002_{\rm stat} \pm 0.005_{\rm sys})$~meV.
	The systematic uncertainty is obtained from the combined contributions of uncertainties in: energy loss in the neon gas ($1.7 \%$ the uncertainty in our proton energy range), beam heating correction ($1.6$\%), energy straggling effect ($1$\%), and  efficiency ($4$\%).
	Our resonance strength is in agreement with the previous value \cite{Rolfs75-NPA} but has an overall uncertainty reduced by a factor of $3$ (from $18\%$ to $5\%$).
	
	Finally, the direct capture component was measured for the first time below the $E\mathrm{_{cm}} = 368.0$~keV resonance, at $E\mathrm{_{cm}}$ = $247.6$, $250.5$, $252.4$, $284.9$, $294.8$, $303.7$, $313.8$, and $362.1$~keV using natural neon gas at a pressure of $2$~mbar, and at $E\mathrm{_{cm}}$ = $380.9$~keV using a pressure of $0.5$~mbar (the latter pressure was chosen to avoid populating the $E\mathrm{_{cm}} = 368.0$~keV resonance).
	For all beam energies investigated, the $\rm DC\rightarrow 2425~keV$ transition occurred in a region of the spectrum affected by the laboratory background and thus with a low signal/noise ratio.
	Hence, we used the secondary transition ($2425~{\rm keV} \rightarrow {\rm GS}$) instead, exploiting the fact that the two $\gamma$ rays (primary and secondary) occur in cascade (i.e., one-to-one correspondence).
	The corresponding $\gamma$-ray line 
	was observed with a statistical uncertainty between $\sim 3$\% at $E\mathrm{_{cm}} = 294.8$~keV and $\sim 30$\% at $E\mathrm{_{cm}} = 252.4$~keV. 
 
\begin{figure}[t!]
		\includegraphics[width=0.5\textwidth]{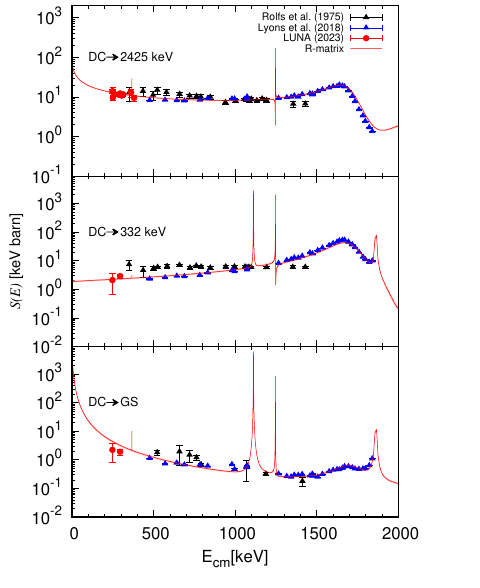}
		\caption{(From top to bottom): LUNA $S$-factor values (red data points) for direct transitions to the second- ($2425$~keV), first- ($332$~keV), and ground states in $^{21}$Na.
			Note that the ${\rm DC}\rightarrow 2425~{\rm keV}$ transition was analyzed using the secondary $\gamma$ rays ($2425~{\rm keV} \rightarrow {\rm GS}$) (see text for details). 
			Error bars for the LUNA data include statistical and systematic uncertainties. 
			The red curve shows our global R-matrix fit to the total $S$-factors, and includes data by Rolfs \textit{et al.} \cite{Rolfs75-NPA} (black data points) and by Lyons \textit{et al.} \cite{Lyons18-PRC} (blue data points). 
		}
		\label{fig_3}
	\end{figure}

 The direct capture energy region explored was affected by beam-induced background caused by $^{14}$N and $^{19}$F contaminants which reduce the signal/noise ratio. Therefore the weakest transitions (${\rm DC} \rightarrow {\rm GS}$ and ${\rm DC} \rightarrow 332$~keV) were observed only at $E\mathrm{_{cm}}$ = $247.6$~keV and $294.8$~keV where the contribution of the beam-induced background did not limit the signal/noise ratio.
	In an extended gas target, fusion reactions take place in the entire target chamber. 
    Therefore, the $S$-factor is determined by the following relationship: 
	\begin{equation}
		Y(E) = \int_{0}^{z\mathrm{_{cal}}} S(E(z))\frac{e^{-2\pi\eta(E(z))}}{E(z)} \rho(z)\widetilde{\eta}(z)\mathrm{d}z,
		\label{eq_4}
	\end{equation}
	\noindent where $Y(E)$ is the experimental yield for each beam energy, $\widetilde{\eta}(z)$ is the efficiency as a function of the position in the chamber, $z\mathrm{_{cal}}$ is the position of the calorimeter surface, and $E(z)$ is the center-of-mass energy along the target path.
 Extracted $S$-factor values for all transitions are shown in Fig.~\ref{fig_3} together with literature data\footnote{The total $S$-factor from Ref. \cite{Lyons18-PRC} has been obtained using the angular coefficient reported in \cite{Lyons18-PRC}}.

 The new LUNA S-factor values for the different transitions are given in the Supplemental Materials.
	Available $S$-factor data, including the present work, were analyzed using the R-matrix formalism with the AZURE2 code to extrapolate down to astrophysical energies \cite{Azuma_2010}.
 Fitting parameters (channel radius, excited states properties, etc.) for the R-matrix analysis were taken from Ref. \cite{Lyons18-PRC}, except the ANC coefficients, which were taken from Ref. \cite{Mukhamedzhanov2006AsymptoticNC} (see details in Supplemental Materials \cite{Supplement}).
	The stated systematic uncertainty of 10$\%$ was used for the data of Lyons \textit{et al.} \cite{Lyons18-PRC}, while a conservative systematic uncertainty of 20$\%$ was assumed for the data of Rolfs \textit{et al.} \cite{Rolfs75-NPA} since no detailed description of the error budget is given in Ref. \cite{Rolfs75-NPA}.
	For the LUNA $S$-factor data, we adopted systematic uncertainty of $\leq 6.6$\%, obtained from the same source of uncertainties reported for the resonant contribution. 
 The largest value (6.6\%) of systematic uncertainties is due to the fluctuation in the beam current.
 The resulting R-matrix fit is shown in Fig.~\ref{fig_3} and the normalization coefficients for direct transitions to the second- ($2425$~keV), first- ($332$~keV), and ground states in $^{21}$Na are 1.038, 0.989 and 1.053, respectively.

Our new data mainly constrain the $S$-factor extrapolations at low energies. 
 The fit was performed in the Bayesian framework using the BRICK package \cite{brick}. We have used a continuous uniform distribution as a prior distribution for all the parameters when no foreknowledge is assumed. Instead, for the ANCs, measured with indirect methods,  we considered a Gaussian distribution.

	Finally, we calculated an updated thermonuclear reaction rate for the $\mathrm{^{20}Ne(p, \gamma)^{21}Na}$ reaction using the results reported in this work. 
	The DC component was taken from the R-matrix fit, whereas the resonant contributions were added following the narrow resonance formalism \cite{Iliadis07-Book}. 
	For the latter, our new values of resonance energy and strength were used. 
	The $E^{\rm }_{cm}$ = $397$~keV, $1247$~keV, $1430$~keV, $1862$~keV resonances reported in Ref. \cite{ILIADIS_2010R} and $E\mathrm{_{cm}}$ = $1113$~keV resonance from Ref. \cite{Christian_2013} were also included.
	
\begin{figure}[ht]
	\includegraphics[width=0.5\textwidth]{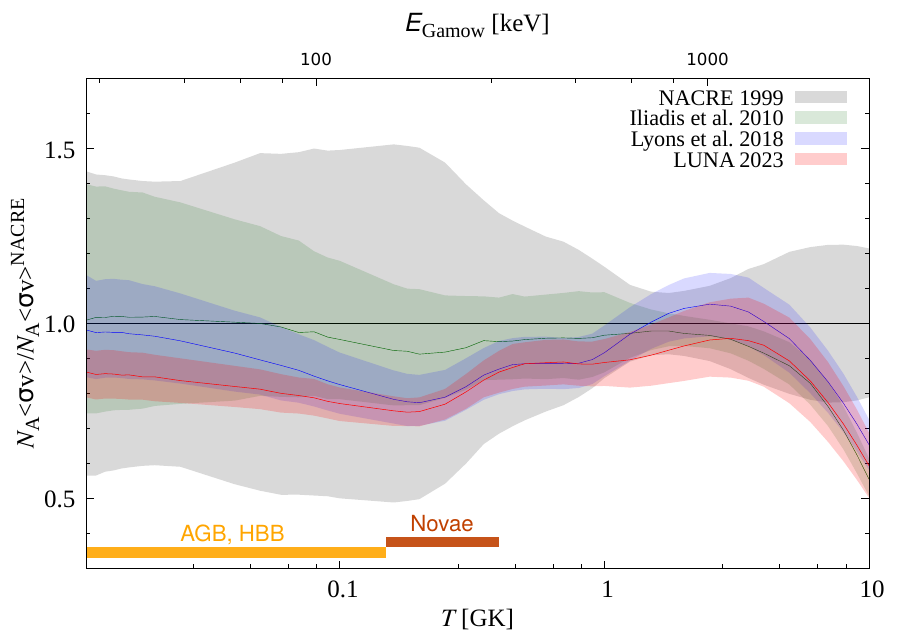}
		\caption{LUNA thermonuclear reaction rate for the $\mathrm{^{20}Ne(p, \gamma)^{21}Na}$ reaction (red) compared with NACRE \cite{NACRE99-NPA} (grey), Iliadis {\it et al.} \cite{ILIADIS2010_II} (green), and Lyons $et$ $al$. \cite{Lyons18-PRC} (blue) rates. The shaded area shows
the $1\mathrm{\sigma}$-uncertainty. The rates are normalized to NACRE. The top x-axis shows the Gamow energy for the given temperature range.}
\label{fig_4}
\end{figure}

The new reaction rate is shown in Fig.~\ref{fig_4}, 
relative to the standard NACRE rate \cite{NACRE99-NPA}. 
The new rate is generally lower than previous rates \cite{NACRE99-NPA,ILIADIS2010_II, Lyons18-PRC}, except for temperatures  $T = 0.2- 1.0$~GK, where it is dominated by the $E^{\rm }_{cm}$ = $368.0$~keV resonance and by the tail of the $E^{\rm }_{cm}$ = $1113$~keV resonance. 
At these temperatures, the LUNA rate is about 3$\%$ higher compared to Ref.~\cite{Lyons18-PRC} (blue), and about 5\% lower compared to Ref.~\cite{ILIADIS2010_II} (green). 

For $T<0.1$ GK (corresponding to $E<$200 keV), the present data points are unique and show that the rate is 20\% lower than NACRE (Fig. \ref{fig_4}).

The new rate can affect mainly two astrophysical scenarios: AGB stars and their HBB phase and classical novae. In the following, we describe the nucleosynthetic impact for both cases.


We performed nucleosynthesis calculations for the TP-AGB phase of stars with an initial mass of $3\, M_{\odot}$,  $4\, M_{\odot}$ and $5\, M_{\odot}$ and low metallicity, $Z=0.0002$.
In the deepest layers of the convective envelope, where the Ne-Na cycle operates, the $5\, M_{\odot}$ model experiences a strong HBB, reaching temperatures up to $\simeq 0.1$~GK. 
To estimate the effect of the $\mathrm{^{20}Ne(p, \gamma)^{21}Na}$ rate in the adopted models, for each of them we perform the calculations using the LUNA rate, the NACRE \cite{NACRE99-NPA} rate as a reference, and also the Iliadis {\it et al.} \cite{ILIADIS2010_II} rate.

For the $5\, M_{\odot}$ model experiencing powerful HBB, our rate reduces the surface abundance of $^{21}$Ne by about 26\% with respect to NACRE\cite{NACRE99-NPA}. $^{23}$Na and $^{22}$Ne are reduced by 10\% and 5\%, respectively. The starting isotope $^{20}$Ne is abundant, and our new rate has no effect on its surface abundance, as expected.
Regarding the other TP-AGB models with initial masses of $3\, M_{\odot}$ and  $4\, M_{\odot}$ the temperature at the base of their convective envelopes remains well below 0.1 GK; hence, the new rate has only a slight effect on the NeNa cycle.
The uncertainty in the nuclear reaction rate is now much smaller compared to other theoretical uncertainties, specifically those related to convective instabilities. As a result, future measurements of element abundance in the atmospheres of AGB stars will better constrain stellar convection theory.

	Subsequently, we have performed 12 dedicated hydrodynamic simulations of oxygen-neon (ONe) novae. We used the spherically-symmetric (1D), implicit, Lagrangian, hydrodynamic SHIVA code, extensively used in the modeling of stellar explosions \cite{Jose_1998,Jose16-Book}.
 
  We adopted a representative case, with an ONe white dwarf accreting solar composition material from a companion, main sequence star, at a rate of 2$\times10^{-10}$ $M_\odot$ per year. The accreted material is assumed to mix with material from the outer layers of the underlying white dwarf to a characteristic level of 50\%. The white dwarf initially has a luminosity of 0.01 times the solar value. Three different values for the white dwarf mass (i.e., 1.15$M_\mathrm{\odot}$, 1.25$M_\mathrm{\odot}$, and 1.35$M_\mathrm{\odot}$) have been adopted to evaluate the impact of the new rate for different thermal histories. For each mass, we have computed 4 hydrodynamic models, identical to one another, except for the prescription adopted for the $^{20}$Ne($p,\gamma$)$^{21}$Na rate: First, we compared our rate with NACRE, to estimate the impact of the changed rate on nova ejecta. Afterwards, we also used the upper and lower limits of our rate to quantify the remaining uncertainties contributed by our new rate to predicted nova yields.

  When using our rate, we find a reduction of up to 23\% in ejected radioactive $^{22}$Na. 
  This isotope is important because the direct detection of its decay in space-based $\gamma$-ray spectrometers such as INTEGRAL or the future COSI would present a ``smoking gun'' for nova nucleosynthesis \cite{Fougeres23-NatComms}. 
  Other isotopes affected by our rate include $^{21,22}$Ne (30\% reduction, important for neon isotopic ratios in grains of possible nova origin \cite{Pepin11-ApJ,Pepin19-MAPS}), and $^{23}$Na, $^{24, 25, 26}$Mg, and $^{26,27}$Al (10-20\% reduction depending on isotope). 
  The comparison of the yields obtained with our upper and lower limits shows a variation of just 1-10\% for key species in the Ne-Si group, meaning our rate provides a firmer basis to characterize nova yields with unprecedented precision. 

	In summary, we reported a new determination of the energy of the $E\mathrm{_{cm}}$ = ($368.0 \pm 0.5$)~keV resonance in $\mathrm{^{20}Ne(p, \gamma)^{21}Na}$, its strength and branching ratios and, for the first time, of the direct capture component at $E_{\rm cm} < 352$~keV. 
	Our resonance strength value has a factor-of-3 lower uncertainty compared to the literature.
	The resonance energy is slightly higher than previously reported, albeit still in agreement within uncertainties. 
	The direct capture cross-section data below $352$~keV were measured with improved systematic ($< 7 \%$) and statistical uncertainties ($3\% - 30\%$). 
	Our new data put a stronger constraint on the extrapolated $S$-factor and on the contribution of the sub-threshold resonance, as reflected in our improved thermonuclear reaction rate. 
Based on dedicated nucleosynthesis calculations, we find that the production of key neon, sodium, and aluminum isotopes is reduced by 5-40\% both in AGB stars and in ONe novae, in particular 20-23\% reduction in nova-produced $^{22}$Na.
  
  \paragraph{Acknowledgments $-$} 
	D. Ciccotti and the technical staff of the LNGS are gratefully acknowledged for their support.
	We acknowledge funding from: INFN,
	the Italian Ministry of Education, University and Research (MIUR) through the ``Dipartimenti di eccellenza'' project ``Physics of the Universe'',
	the European Union (ERC-CoG STARKEY, no. 615604; ERC-StG SHADES, no. 852016; and ChETEC-INFRA, no. 101008324),
	Deut\-sche For\-schungs\-ge\-mein\-schaft (DFG, BE~4100-4/1),
	the Helm\-holtz Association (ERC-RA-0016),
	the Hungarian National Research, Development and Innovation Office (NKFIH K134197),
	the European Collaboration for Science and Technology (COST Action ChETEC, CA16117).
	T.S. acknowledges support from the J\'anos Bolyai research fellowship of the Hungarian Academy of Sciences.
    JJ acknowledges support by the Spanish MINECO grant PID2020-117252GB-I00, and by the AGAUR/Generalitat de Catalunya grant SGR-386/2021
	M.A., C.G.B, T.D., and R.S.S. acknowledge funding from STFC (grant ST/P004008/1).
	R.J. deBoer is also gratefully acknowledged for his support in the use of the AZURE2 software for R-matrix fitting.
	For the purpose of open access,  authors have applied a Creative Commons Attribution (CC BY) licence to any Author Accepted Manuscript version arising from this submission.
\end{document}